# Reactive additive capillary stamping with double network hydrogel-derived aerogel stamps under solvothermal conditions

Fatih Alarslan,[‡,§] Martin Frosinn,[‡,§] Kevin Ruwisch,[$] Jannis Thien,[$] Tim Jähnichen,[&] Louisa Eckert,[&] Jonas Klein,[§] Markus Haase,[§] Dirk Enke,[&] Joachim Wollschläger,[$] Uwe Beginn,[§] Martin Steinhart[*,§]

[§] Institut für Chemie neuer Materialien and CellNanOs, Universität Osnabrück, Barbarastr. 7, 49076 Osnabrück, Germany

[$] Department of Physics, Universität Osnabrück, Barbarastr. 7, 49076 Osnabrück, Germany

[&] Institute of Chemical Technology, Universität Leipzig, Linnéstraße 3, 04103 Leipzig, Germany

[‡] F. A. and M. F. contributed equally to this manuscript

## Abstract

Integration of solvothermal reaction products into complex thin-layer architectures is frequently achieved by combinations of layer transfer and subtractive lithography, whereas direct additive substrate patterning with solvothermal reaction products has remained challenging. We report reactive additive capillary stamping under solvothermal conditions as a parallel contact-lithographic access to patterns of solvothermal reaction products in thin-layer configurations. To this end, corresponding precursor inks are infiltrated into mechanically robust mesoporous aerogel stamps derived from double-network hydrogels (DNHGs). The stamp is then brought into contact with a substrate to be patterned under solvothermal reaction conditions inside an autoclave. The precursor ink forms liquid bridges between the topographic surface pattern of the stamp and the substrate. Evaporation-driven enrichment of the precursors in these liquid bridges along with their liquid-bridge-guided conversion into the solvothermal reaction products yields large-area submicron patterns of the solvothermal reaction products replicating the stamp topography. As example, we prepared thin hybrid films, which contained ordered monolayers of superparamagnetic submicron nickel ferrite dots prepared by solvothermal capillary stamping surrounded by nickel electrodeposited in a second, orthogonal substrate functionalization step. The submicron nickel ferrite dots acted as magnetic hardener halving the remanence of the ferromagnetic nickel layer. In this way, thin-layer electromechanical systems, transformers and positioning systems may be customized.

# Introduction

Solvothermal syntheses carried out in autoclaves at temperatures exceeding the boiling point of the used solvents may yield reaction products difficult to attain under ambient conditions.[1, 2] Alternative syntheses typically require much higher reaction temperatures or involve high-temperature calcination steps. Thus, metal-organic frameworks,[3] hierarchical nanostructures for water splitting,[4] multi-dimensional noble-metal-based catalysts for electrocatalysis,[5] semiconductors such as zinc oxide,[6] ferrites,[7] perovskite oxides,[8] $LiFePO_4$ for Li-ion batteries[9] as well as two-dimensional transition metal carbide and nitride hybrids for catalytic energy storage and conversion[10] were obtained by solvothermal syntheses. The manufacturing of functional components containing patterned thin layers of solvothermal reaction products has remained challenging. It is conceivable to use continuous thin films of solvothermal reaction products[11] as starting material. Potential approaches to generate patterned films of solvothermal reaction products on receiving substrates then may comprise standard lithography under ambient conditions,[12] combinations of layer transfer by wafer bonding or ion slicing and subtractive lithography as well as laser-induced forward transfer (LIFT).[13] It is obvious that the direct lithographic deposition of patterns of solvothermal reaction products on receiving substrates would be more efficient. However, additive lithographic methods including inkjet and aerosol jet printing[14] as well as classical soft lithography[15, 16] and polymer pen lithography[17-19] typically only enable the deposition of precursors; their conversion into the solvothermal reaction products commonly comes along with the destruction of the lithographically generated precursor patterns, for example, because the solvent used for the conversion dissolves the precursors.

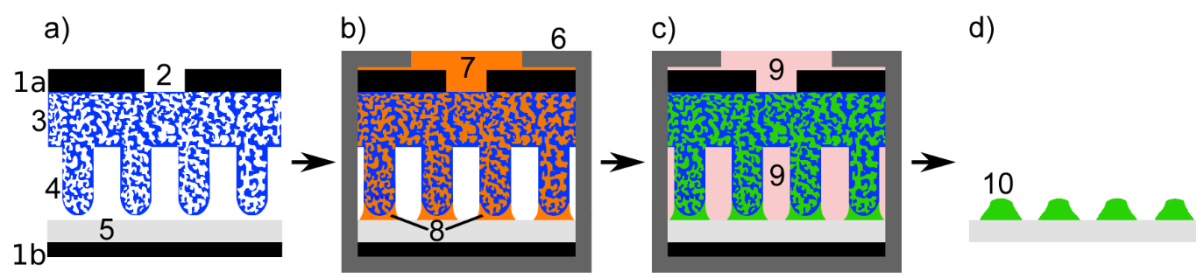

**Figure 1.** Additive solvothermal capillary stamping. a) A stamping device (**1**) consists of an upper part (**1a**) with an ink supply hole (**2**) and a lower part (**1b**). A DNHG-derived aerogel stamp (**3**) with contact elements (**4**) is attached to upper part (**1a**) and approached to a substrate (**5**) attached to lower part (**1b**). b) Stamping device (**1**) is inserted into PTFE vessel (**6**). Precursor solution (**7**) deposited on the surface of upper part (**1a**) flows through hole (**2**) into DNHG-derived aerogel stamp (**3**) and forms liquid bridges (**8**) between contact elements (**4**) and substrate (**5**). c) PTFE vessel (**6**) is then located in a sealed steel autoclave (**11**) (not shown, cf. Figure 2). Under solvothermal conditions, an ethanol-rich gas phase (**9**) fills PTFE vessel (**6**). The nonvolatile precursors of the solvothermal reaction product (**10**) and eventually solvothermal reaction product (**10**) itself enrich in the liquid bridges (**8**). d) After removal from stamping device (**1**) and detachment of DNHG-derived aerogel stamp (**3**), substrate (**5**) is modified by patterns of solvothermal reaction product (**10**), here arrays of submicron nickel ferrite dots.

The direct deposition of patterns of solvothermal reaction products may be achieved by the coupling of solvothermal syntheses with *in situ* additive substrate patterning. It was shown that porous stamps[20-25] are particularly suitable for parallel additive substrate patterning even if diluted precursor solutions are used as inks because the pore systems of the stamps act as ink reservoirs. Evaporation of the volatile ink components (typically the solvents) drags more and more of the non-volatile ink components, such as precursors of solvothermal reaction products, into the liquid ink bridges formed between the contact elements of the stamp and the substrate to be patterned. The enrichment of the non-volatile ink components in the liquid bridges is a crucial aspect of additive pattern formation with porous stamps.[22] However, the porous stamps so far available are not compatible with solvothermal processes; polymeric stamps[20, 21, 23] are too deformable, whereas silica[22, 24] and phenolic resin stamps[25] are too brittle. In general, reactive additive lithography has, so far, predominantly been conducted with solid stamps under ambient pressure and temperature.[26]

Here we report the direct contact-lithographic generation of thin patterned layers of solvothermal reaction products by additive solvothermal capillary stamping with aerogel stamps derived from double network hydrogels (DNHGs)[27, 28] (Figure 1). DNHG-derived aerogel stamps contain continuous mesopore systems and exhibit excellent mechanical strength because their scaffold is a combination of a hard but brittle network and a soft and ductile network. Hence, the DNHG-derived aerogel stamps can be used for parallel additive contact lithography under solvothermal reaction conditions and enable the generation of patterns of solvothermal reaction products by stamp-guided solvothermal conversion of their precursors. As example, we demonstrate the preparation of thin nickel ferrite-nickel ($NiFe_2O_3$-Ni) hybrid layers consisting of a regular monolayer of submicron nickel ferrite dots embedded in a continuous nickel film. Nickel ferrite nanostructures and nanocomposites[29] have, for example, attracted significant interest for magnetic hyperthermia,[30, 31] as well as for catalytic and pseudocapacitive energy storage.[32] Capillary stamping with DNHG-derived aerogel stamps combined with a solvothermal synthesis based on ethanolic solutions of iron(III)-acetylacetonate ($Fe(C_5H_7O_2)_3$) and nickel(II)-acetylacetonate ($C_{10}H_{14}NiO_4$)[33, 34] yielded arrays of submicron nickel ferrite dots on indium tin oxide (ITO) substrates. Then, the ITO substrates functionalized with ordered monolayers of submicron nickel ferrite dots were orthogonally functionalized[35] with metallic nickel by electrodeposition. The superparamagnetic submicron nickel ferrite dots embedded into the thin ferromagnetic nickel film halved the magnetic remanence of the latter, while the saturation value of the magnetic moment per area remained by and large unaffected.

## Materials and Methods

### Chemicals and materials

Acrylamide (AAm) (>99%) and zirconium(IV)-chloride (>98%) were purchased from Merck. Potassium peroxodisulfate (PPS) (>99%) and titanium(IV)-chloride (>99%) were purchased from Fluka. *N,N'*-Methylenebis(acrylamide) (MBA) (99%), *p*-toluenesulfonyl chloride (>98%), nickel(II)-acetylacetonate (95%), iron(III)-acetylacetonate (97%), nickel sulfate, sodium chloride, boric acid and potassium dodecyl sulfonate were purchased from Sigma-Aldrich. *N,N,N',N'*-tetramethylethylenediamine (TMEDA) (99 %) and tetraethoxysilane (98 %) were purchased from Alfa Aesar. Ethylene glycol (99.8 %) was purchased from VWR Chemicals. Perfluorodecyltrichlorosilane (FDTS; $C_{10}H_4Cl_3F_{17}Si$) was purchased from abcr GmbH (Karlsruhe). Synthetic hectorite $Na^+_{0.7}[(Si_8Mg_{5.5}Li_{0.3})O_{20}(OH)_4]^{-0.7}$ (Laponite RD) was purchased from GMW (Vilsheim). Macroporous silicon[36, 37] with macropores having a diameter of 1 µm and a depth of 730 nm arranged in a hexagonal array with a lattice constant of 1.5 µm was purchased from Smart Membranes (Halle an der Saale). Indium tin oxide substrates $(In_2O_3)_{0.9} \cdot (SnO_2)_{0.1}$ (resistance 8-12 Ω sq$^{-1}$, thickness 1200–1600 Å) were purchased from Sigma Aldrich. All water used was fully deionized and further purified using a Sartorius stedim/ arium 611 UV device. If not stated otherwise, chemicals were used as received.

### Fabrication of DNHG-derived aerogel stamps

**Preparation of synthetic hectorite/zirconylchloride octahydrate solutions.** To prepare synthetic hectorite solutions, 10 g synthetic hectorite was dissolved in 1 L deionized water. Synthetic hectorite solutions show a complex phase evolution starting with the swelling of stacked hectorite platelets, which is followed by exfoliation and the formation of house-of-cards structures, in which the positive charged sides of the platelets interact with the negative charged faces of surrounding platelets (Figure S1).[38] Therefore, the properties of freshly prepared synthetic hectorite solutions slightly differ from those of aged synthetic hectorite solutions. Thus, all synthetic hectorite solutions used in this work were aged for at least one week. Zirconylchloride octahydrate was prepared by hydrolyzing zirconium chloride in deionized water, followed by evaporation of excess water under reduced pressure until no further mass loss was observed. 100 g of zirconylchloride octahydrate was added to the turbid synthetic hectorite solutions, which were then stirred for one week so that transparent synthetic hectorite/zirconylchloride octahydrate solutions were obtained.

**Preparation of tetrakis-(2-hydroxyethoxy)-silane solutions in ethylene glycol.** Ethylene glycol (300 g, 4.83 mol), tetraethoxyorthosilane (250 g, 1.20 mol) and *p*-toluenesulfonyl chloride (40 mg, 0.21 mmol) were filled into a 1 L round-bottom flask. Vigorous stirring for 60 minutes at 70°C under ambient pressure yielded a homogenous colorless solution. 200 g of the reaction product ethanol was removed from the sol solution by distillation with a rotary evaporator applying a reduced pressure of 300 mbar for two hours and subsequently a reduced pressure of 90 mbar. Thus, a clear, colorless and moisture-sensitive solution containing tetrakis-(2-hydroxyethoxy)-silane as well as oligomeric or polymeric silane species derived from tetrakis-(2-hydroxyethoxy)-silane in ethylene glycol was obtained. For characterization by NMR spectroscopy, 10 to 15 mg of the solution was mixed with 0.5 ml deuterized DMSO. NMR measurements were performed at 30 °C using a Bruker Avance III spectrometer at 500 MHz ($^1$H), 125 MHz ($^{13}$C) or 100 MHz ($^{29}$Si).

$^1$H-NMR (DMSO-d6 δ/ppm) (Figure S2): 1.056 (t, J= 7.05, **CH$_3$**), 1.156 (m, **CH$_3$**), 3.395 (s, **CH$_2$**), 3.443 (m, **CH$_2$**), 3.490 (s, br, **CH$_2$**), 3.749 (s, br. **CH$_2$**), 4.3960 (s, **OH**)

$^{13}$C-NMR (DMSO-d6 δ/ppm) (Figure S3): 18.473 (s, **CH$_3$**), 56.044 (s, **CH$_2$**), 61.574 (m, br, **CH$_2$**), 62.797 (s, **CH$_2$**), 64.854 (s, br, **CH$_2$**)

$^{29}$Si-NMR (DMSO-d6 δ/ppm) (Figure S4): -108.249 (s, br, **SiO$_2$**)

**Preparation of the sol solution for the DNHG synthesis.**
8 g tetrakis-(2-hydroxyethoxy)-silane solution was added to 10 g synthetic hectorite/zirconylchloride octahydrate solution in a 50 ml round-bottom flask under constant stirring. Subsequently, 6 g AAm (84.41 mmol) and 2.5 mg MBA (0.0162 mmol) were added. The flask was placed in an ice bath after a clear solution was obtained. A mixture of 2 µl TMEDA (1.55 mg, 0.0133 mmol) in 0.5 ml H$_2$O was added and the flask was closed with a septum cap. A separate flask with a septum containing 4 mg PPS (0.0148 mmol) and 1 ml H$_2$O was placed in the ice bath as well. Both solutions were suffused with nitrogen for 30 minutes. Under exclusion of oxygen, both solutions were combined and mixed thoroughly. The combined solutions were subjected to three vacuum-ultrasonication cycles under cooling to 0°C using ice baths. Each cycle comprised the application of a vacuum for 30 seconds using a diaphragm pump followed by sonication for 30 seconds, while the vacuum was maintained. The obtained sol solution was kept at 0°C under nitrogen atmosphere and used within 5 minutes.

**Preparation of DNHG-derived aerogel stamps.** The surface of macroporous silicon was treated with oxygen plasma for 10 minutes and then coated with FDTS by chemical vapor deposition at 100 °C for 10 h following procedures reported elsewhere.[39] About 15 mL of the sol solution was poured onto FDTS-coated macroporous silicon pieces extending 1 cm$^2$ located in glass vials under exclusion of oxygen under nitrogen atmosphere. The vials were closed airtight and gelation as well as aging was allowed to take place at 25°C for at least one week. The aged cylindrical DNHG monoliths with a diameter of 24 mm, a height of ~22 mm and a mass of ~16 g having a topographically patterned contact surface were detached from the macroporous silicon under ambient conditions while still being wet. Then, each aged hydrogel monolith was treated with 500 ml methanol for 120 h, 500 ml THF for 120 h and 500 ml *n*- hexane for 120 h in a Soxhlet extractor. This procedure resulted in shrinkage to 78% of the initial volume of the water-filled DNHG. Evaporation of the *n*-hexane at 20 °C in a vacuum of ~0.001 mPa did not lead to further shrinkage. Subsequently, the macroporous silicon template was removed from the gel. As a result, DNHG-derived aerogel stamps (**3**) (cf. Figure 1) were obtained, which had contact surfaces extending 1 cm$^2$ topographically patterned with contact elements (**4**). The contact elements (**4**) had a diameter of 900 nm and a height of 600 nm. Prior to further use, the height of the DNHG-derived aerogel stamps (**3**) was reduced to ~1.3 cm by sawing. The remaining surfaces except the topographically patterned contact surface with the contact elements (**4**) were ground.

**Characterization of the DNHG-derived aerogel**

Nitrogen sorption measurements on DNHG-derived aerogel samples were performed with a device Autosorb from Quantachrome at 77 K. Before any measurement, about 100 mg sample material was degassed at 373 K for 10 h in an ultrahigh vacuum. The specific surface area was calculated using the BET method in a relative pressure range $p/p_0$ = 0.05 - 0.30. The total pore volume was calculated at $p/p_0$ = 0.995. For the calculation of the pore size distribution, the BJH method was applied to the desorption branch of the isotherm. For the analyses, the program ASiQwin (Quantachrome Instruments) was used. Mercury intrusion measurements applied to determine the porosity of the samples were carried out with a Pascal 440 device (Porotec). Intrusion measurements were performed at 297 K up to 400 MPa. The contact angle of mercury was set to 140° and the surface tension to 0.48 N m$^{-1}$. Compressive stress tests on 5 DNHG-derived aerogel specimens extending ~10 mm * ~10 mm * ~5 mm were conducted with a PCE-MTS 500 test stand equipped with a PCE-DFG N 5K force gauge and two parallel steel plates. The compressions of the samples were calculated from the rate of compression (0.43 mm/s with 50 data points per second).

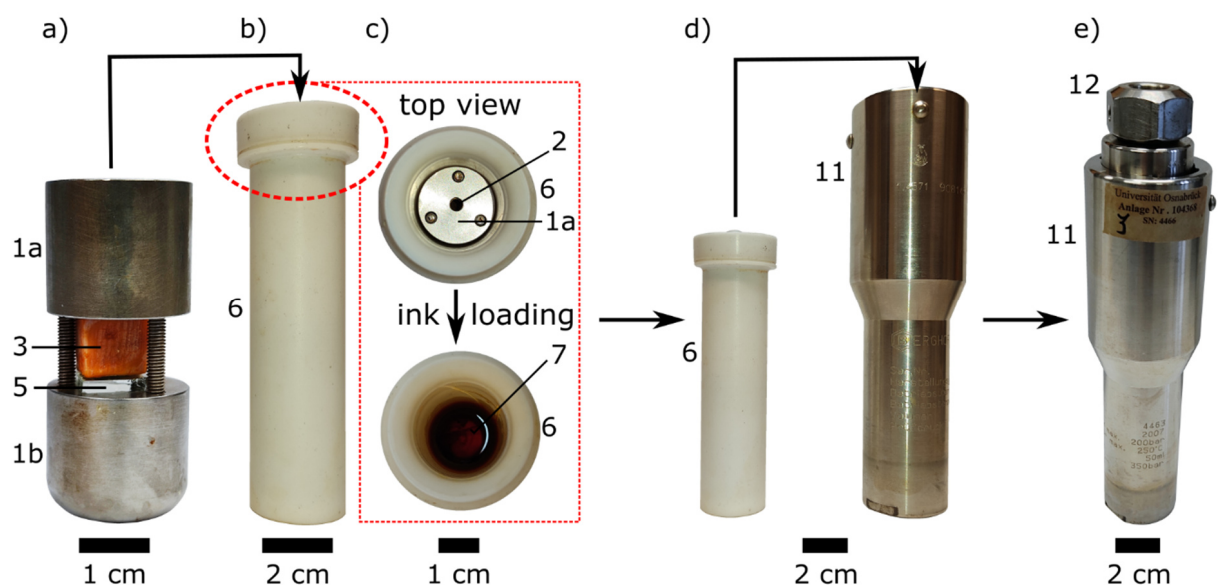

**Figure 2.** Setup for additive solvothermal capillary stamping. a) Stamping device (**1**) consists of an upper part (**1a**), on which a DNHG-derived aerogel stamp (**3**) is mounted, and a lower part (**1b**), on which substrate (**5**) (here: FDTS-coated ITO) is placed. Upper part (**1a**) and lower part (**1b**) are then connected by pins preventing lateral displacement. b) Stamping device (**1**) including the DNHG-derived aerogel stamp (**3**) and substrate (**5**) is transferred into PTFE vessel (**6**). c) Upper part (**1a**) of stamping device (**1**) containing hole (**2**) is covered with precursor solution (**7**) while located in PTFE vessel (**6**). Precursor solution (**7**) flows through hole (**2**) into the DNHG-derived aerogel stamp (**3**). d) PTFE vessel (**6**) including stamping device (**1**), DNHG-derived aerogel stamp (**3**), substrate (**5**) and precursor solution (**7**) is inserted into steel autoclave (**11**). e) Steel autoclave (**11**) is then sealed with lid (**12**).

**Additive solvothermal capillary stamping**

The stamping device (**1**) consisted of an upper part (**1a**) and a lower part (**1b**). The upper part (**1a**) consisted of a stainless-steel cylinder with a diameter and a height of 2 cm. The steel cylinder contained three peripheral holes parallel to the cylinder axis with a diameter of 3 mm forming an equilateral triangle with an edge length of 1.4 cm as well as a central hole (**2**) parallel to the cylinder axis with a diameter of 4 mm. A DNHG-derived aerogel stamp (**3**) was glued onto the upper part (**1a**) with double-sided adhesive tape in such a way that the contact elements (**4**) of the DNHG-derived aerogel stamp (**3**) pointing away from upper part (**1a**) where exposed (cf. Figure 1). The lower part (**1b**) of the stamping device (**1**) was an exact copy of the upper part (**1a**) except that it did not contain a central hole (**2**). The ITO substrates (**5**) were coated with FDTS applying the same procedure reported elsewhere[39] as in the case of macroporous silicon. The contact elements (**4**) of the DNHG-derived aerogel stamp (**3**) glued on the upper part (**1a**) were brought into contact with an FDTS-coated ITO substrate (**5**) placed on the lower part (**1b**) (Figure 1a). The upper part (**1a**) was then fixed on the lower part (**1b**) by inserting pins into the three peripheral holes of upper part (**1a**) and the corresponding counterpart holes in the lower part (**1b**) (Figure 2a). The pins prevented lateral displacement of the upper part (**1a**) and the lower part (**1b**) with respect to each other. Furthermore, the pins prevented lateral displacement of the FDTS-coated ITO substrate (**5**), which was not fixated otherwise on the lower part (**1b**). The contact pressure exerted by the DNHG-derived aerogel stamp (**3**) on the FDTS-coated ITO substrate (**5**) amounted to 3.9 kN/m$^2$ and originated from the mass of the upper part (**1a**) of 40 g. The assembled stamping device (**1**) including the DNHG-derived aerogel stamp (**3**) and the FDTS-coated ITO substrate (**5**) was then inserted into a cylindrical PTFE vessel (**6**) with a height of 12 cm, a diameter of 3.8 cm and a total volume of 40 mL (Figure 2b). The upper part (**1a**) of stamping device (**1**) fitted into PTFE vessel (**6**) in such a way that the rim of the upper part (**1a**) was in close, self-sealing contact with the wall of the PTFE vessel (**6**). Nickel ferrite precursor solution (**7**) was prepared by dissolving 54.8 mg (0.2 mmol) nickel(II)-acetylacetonate and 150.68 mg (0.4 mmol) iron(III)-acetylacetonate in 50 mL ethanol. 6 mL of precursor solution (**7**) was deposited on the surface of the upper part (**1a**) of the stamping device (**1**) located in PTFE vessel (**6**), flowed through the central hole (**2**) in the upper part (**1a**) and infiltrated the DNHG-derived aerogel stamp (**3**) (Figures 1b and 2c). Then, the PTFE vessel (**6**) containing the assembled stamping device (**1**) loaded with precursor solution (**7**) was inserted into a steel autoclave (**11**), which was then sealed with a steel lid (**12**) (Figures 1c and 2d). The solvothermal reaction yielding arrays of submicron nickel ferrite dots (**10**) was carried out in a pressure digestion system (Berghof digestec DAB-2) at 413 K (140°C) for 48 hours. After completion of the synthesis and disassembly of the stamping device (**1**), the DNHG-derived aerogel stamp (**3**) was detached from the substrate (**5**) patterned with submicron nickel ferrite dots (**10**) (Figure 1d), which was washed with ethanol and dried at 40°C for 1 h. The volume of the autoclave (**11**) equipped with PTFE vessel (**6**) and stamping device (**1**) including a DNHG-derived aerogel stamp (**3**) and an ITO substrate (**5**) available to a fluid phase in its interior was determined by differential weighing before and after filling with water and amounted to ~29.15 mL.

**Electrodeposition of Nickel**

For the electrodeposition of nickel, ITO substrates (**5**), which were either FDTS-coated or functionalized with submicron nickel ferrite dots (**10**), were placed in an electrochemical cell. We used an aqueous plating solution containing 1 mol/L NiSO$_4$*6H$_2$O and 0.1 mol/L of the other components H$_3$BO$_3$, NaCl and NaC$_{12}$H$_{25}$SO$_4$.[40] Thus, 0.536 g NiSO$_4$*6H$_2$O (0.02 mol), 0.124 g H$_3$BO$_3$ (0.002 mol), 0.117 g NaCl (0.002 mol) and 0.557 g NaC$_{12}$H$_{25}$SO$_4$ (0.002 mol) dissolved in 20 ml water were deposited into the electrochemical cell. The ITO substrate (**5**) acted as working electrode, a platinum wire as counter electrode and an Ag/AgCl electrode as reference electrode. The electrodeposition was carried out at a constant current of 1 mA/cm$^2$ for 10 min using a potentiostat Interface 1000 (Gamry). The thickness of the nickel films was determined by scanning electron microscopy investigation of a cross-sectional specimen (Figure S5).

**Contact angle measurements**

Contact angles of the nickel ferrite precursor solution (0.2 mmol nickel(II)-acetylacetonate and 0.4 mmol iron(III)-acetylacetonate dissolved in 50 mL ethanol) on FDTS-modified ITO substrates were measured by the sessile drop method with a DSA100 drop shape analyzer at 22°C and a relative humidity of 33 %. Overall 6 measurements at different positions were carried out.

**Microscopy and energy-dispersive X-ray spectroscopy**

Prior to scanning electron microscopy (SEM) imaging, the samples were dried overnight at 40 °C in air and then sputter-coated 2-3 times for 15 s with platinum/iridium alloy using an EMITECH K575X sputter coater. SEM images were taken with a Zeiss Auriga device equipped with a field emission cathode and a Gemini column with a working distance of 5 mm applying an acceleration voltage of 3 kV. For image detection an InLens detector was used. For energy-dispersive X-ray (EDX) spectroscopy mappings a X-Max 80 mm$^2$ silicon drift detector (Oxford Instruments) was used. The EDX maps extending 1024x788 pixels were recorded with a pixel dwell time of 5000 µs and a frame live time of 1.05 h. Atomic force microscopy (AFM) topography images were recorded with an NTEGRA microscope (NT-MDT) in the tapping mode using HQ:NSC16/AL BS cantilevers from µmasch with a resonance frequency of 170 - 210 kHz and a force constant of 30 - 70 N/m.

**XPS**

XPS measurements were carried out under ultra-high vacuum using an ESCA system Phi 5000 VersaProbe III with a base pressure of 1•10$^{-9}$ mbar equipped with a monochromatized aluminum anode (Kα = 1486.6 eV) and a 32-channel electrostatic hemispherical electronic analyzer. An ion gun and an electron gun were used to prevent sample charging. A take-off angle of 45° was used. The XP spectra were calibrated using the carbon C 1s peak at 284.5 eV.[41]

**X-ray diffraction (XRD) of submicron nickel ferrite dots**

XRD scans were carried out in theta-theta geometry with Cu Kα radiation using an X'Pert Pro MPD diffractometer (PANalytical) equipped with a rotating sample plate and a PixCel 1D detector. The diffractometer was operated at a voltage of 40 kV and a current of 30 mA. For the XRD measurements, ~2 mg submicron nickel ferrite dots (**10**) were carefully scraped off from several ITO substrates (**5**) with scalpels.

**Magnetometry**

Magnetization curves of ITO substrates (**5**) with an area of 1 cm$^2$ functionalized with arrays of submicron nickel ferrite dots (**10**), continuous nickel layers with a thickness of ~100 nm or nickel ferrite-nickel hybrid films (**13**) (cf. Figure 7 below) were measured at 300 K by varying the magnetic field from -1 T to 1 T using a vibrating sample magnetometer (LakeShore, Model 7404). Magnetization loops were corrected by subtracting the diamagnetic contribution from the substrate.

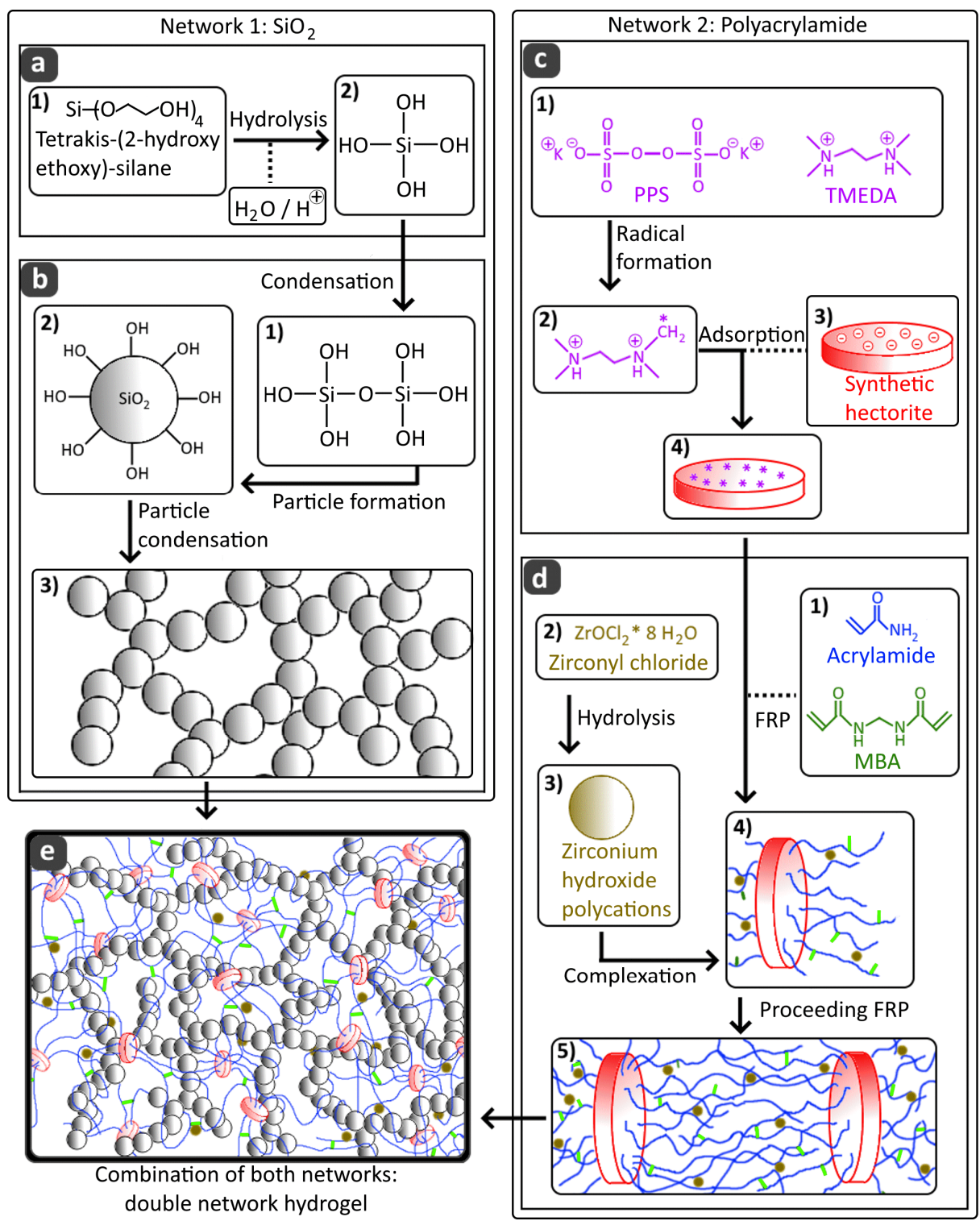

**Figure 3.** Overview of the general synthesis procedure of DNHGs. a), b) Conversion of tetrakis-(2-hydroxy-ethoxy)-silane to a hard SiO$_2$ particle network a) Hydrolysis of tetrakis-(2-hydroxy-ethoxy)-silane yielding hydroxysilane species, which b) form a SiO$_2$ particle network by polycondensation. c), d) Generation of the soft polyacrylamide network. c) Generation of initiation sites (pink stars in c4) for d) the free radical polymerization (FRP) of polyacrylamide on synthetic hectorite platelets. The formed polyacrylamide chains (blue) are crosslinked by the synthetic hectorite platelets, methylene-bis-acrylamide (green) and zirconium hydroxide polycations (brown) (panel c5). e) Resulting DNHG structure.

# Results and discussion

## Synthesis of DNHGs

The general procedure for the DNHG synthesis involves two orthogonal polymerizations simultaneously carried out in the same flask (Figure 3). The first polymerization is a polycondensation reaction yielding a hard silica network (Figure 3a,b). The second polymerization is a free radical polymerization (FRP) yielding a soft polyacrylamide network (Figure 3c,d). Two additional components, zirconyl chloride octahydrate and synthetic hectorite platelets (Figure S1), are active in both reactions,[42, 43] as detailed below.

The starting point for the formation of the hard silica network is the preparation of a solution of tetrakis-(2-hydroxyethoxy)-silane in ethylene glycol (Figure 3a1). This silane, which can only be synthesized directly in the solvent and reactant ethylene glycol, cannot be isolated from the reaction mixture. The removal of excess ethylene glycol would shift the reaction equilibrium to higher degrees of polymerization and crosslinking because tetrakis-(2-hydroxyethoxy)-silane molecules may condense under release of an ethylene glycol molecule. Even in the presence of excess ethylene glycol, tetrakis-(2-hydroxyethoxy)-silane forms to some extent oligomers or polymers, as apparent from the $^1$H (Figure S2), $^{13}$C (Figure S3), and $^{29}$Si (Figure S4) NMR spectra of the reaction mixture. Apart from $^1$H signals at 3.394 ppm and 4.396 ppm as well as the $^{13}$C signal at 62.797 ppm originating from ethylene glycol, the NMR spectra contain only broad signals with widths characteristic of oligomers and polymers. Diffusion-ordered NMR spectroscopy (DOSY) (Figure S6) confirmed that the narrow signals ascribed to ethylene glycol are associated with large diffusion coefficients, whereas the broad signals ascribed to tetrakis-(2-hydroxyethoxy)-silane condensation products are associated with a broad distribution of lower diffusion coefficients. When the tetrakis-(2-hydroxyethoxy)-silane/ethylene glycol mixture is brought into contact with water, hydrolysis of the silane species produces metastable hydroxysilane species (Figure 3a2), which further oligomerize and polymerize by condensation reactions (Figure 3b1). After formation of a viscous sol presumably composed of small silica particles and branched polymeric silica structures (Figure 3b2), gelation proceeds by further condensation reactions. Zirconylchloride is strongly acidic and thereby catalyzes hydrolysis and condensation of the silicon dioxide network. The resulting clear, colorless, stiff and brittle silica network consists of a rigid network of covalently bond silica particles with diameters of a few nm (Figure 3b3).[44]

The formation of the soft polyacrylamide network is initiated by the redox initiator system potassium peroxodisulfate/tetramethylethylenediamine (PPS/TMEDA) (Figure 3c1).[45] Methylene-bis-acrylamide (MBA) crosslinks polyacrylamide chains covalently, while zirconium hydroxide polycations formed by the hydrolysis of zirconyl chloride octahydrate noncovalently crosslink polyacrylamide chains.[46] As a result, stretchable polyacrylamide networks are obtained. To understand the impact of the crosslinking components in the synthesis of the soft polyacrylamide networks of the DNHGs, we measured the tensile strengths and the elongations at break of polyacrylamide hydrogels obtained by systematically varying the contents of either MBA or zirconyl chloride octahydrate in the reaction mixtures used for their syntheses (Supporting Text 1 and Figure S7). The tensile strength of the polyacrylamide hydrogels at first increases with increasing amounts of MBA, passes a maximum and decreases again for high MBA contents (Figure S8), likely because of increasing degrees of inhomogeneity related to the hydrophobic nature of MBA.[47] The elongation at break decreases with increasing amounts of MBA and, therefore, increasing degrees of crosslinking (Figure S9). The tensile strength of the polyacrylamide gels increases (Figure S10) and the elongation at break decreases (Figure S11) with increasing amounts of zirconyl chloride octahydrate in the reaction mixtures related to the increasing degree of non-covalent crosslinking of the polyacrylamide chains by zirconium hydroxide polycations.

Synthetic hectorite platelets are known to interact with the PPS/TMEDA initiator system.[43] Namely, the N,N'-tetramethyethylenediammonium radical formed by this initiator system (Figure 3c2) adsorbs to the negatively charged surface of the synthetic hectorite platelets (Figure 3c3) and generates sites for the initiation of polyacrylamide chain growth on the surfaces of synthetic hectorite platelets (Figure 3c4). Starting from these sites, acrylamide and MBA (Figure 3d1) are copolymerized by FRP. As mentioned above, zirconyl chloride octahydrate (Figure 3d2) undergoes hydrolysis to form zirconium hydroxide polycations (Figure 3d3), which crosslink polyacrylamide chains (Figure 3d4).[46] Crosslinking by MBA as well as radical recombination of the active ends of growing polyacrylamide chains having their starting points adsorbed on synthetic hectorite platelets are further crosslinking modes (Figure 3d5). Preliminary experiments revealed that the addition of synthetic hectorite increases both the tensile strength (Figure S8) and the elongation at break (Figure S9) of polyacrylamide hydrogels, indicating that the synthetic hectorite reduces the amount of free polymer chain ends[43] and contributes crosslinking points only at the starting points of the polyacrylamide chains.

The hard silica network and the soft polyacrylamide network eventually form a DNHG without apparent macroscopic phase separation (Figure 3e). Preliminary experiments revealed that increasing the amount of tetrakis-(2-hydroxyethoxy)-silane in reaction mixtures used to synthesize DNHGs results at first in an increase in the tensile strength and the elongation at break until maximum values are reached. Further increases in the amount of tetrakis-(2-hydroxyethoxy)-silane result in decreases in the tensile strength and the elongation at break (Figures S12 and S13). This outcome can be rationalized by the assumption that, as long as the hard and brittle network has a sufficiently lower mass fraction then the soft and ductile network, upon deformation the brittle network may break into smaller clusters, which act as sliding crosslinks dissipating energy.[28] The DNHG formulation used here is optimized so as to convert the resulting DNHG into an aerogel with high mechanical robustness. An important structural feature of the obtained DNHG is the inter-crosslinking between the polyacrylamide and silica networks by the zirconium-containing species and the synthetic hectorite. Zirconium forms mineral zirconium(IV)silicate with silicon, while attractive interactions exist between zirconium ions and (poly)acrylamide.[46] Co-condensation of the synthetic hectorite platelets with the silica species and synthetic hectorite-zirconium ion interactions further strengthen the interactions between the polyacrylamide and silica networks. The obtained colorless DNHG shows high mechanical strength and is and transparent to slightly opaque.

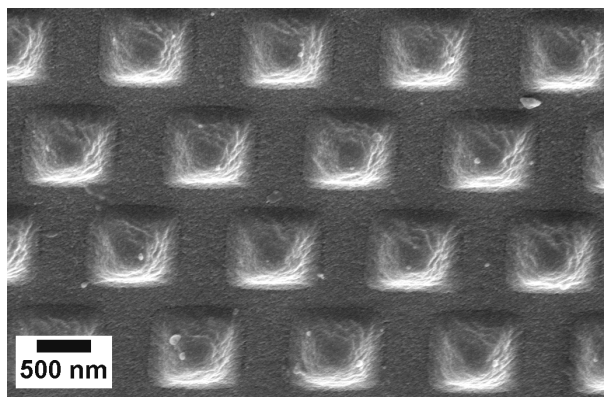

a)

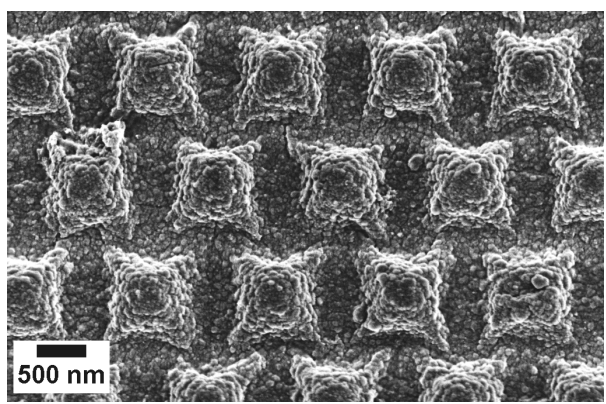

b)

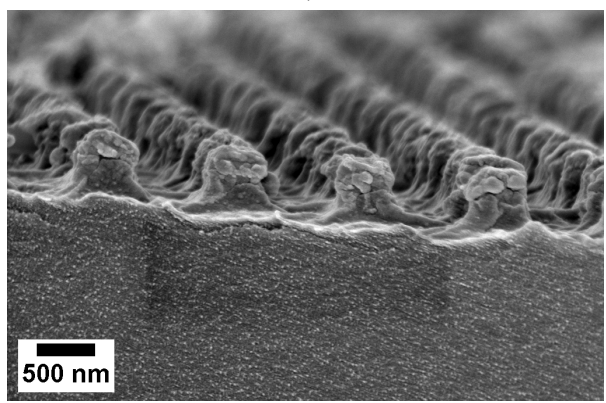

c)

**Figure 4.** SEM images of the contact surfaces of DNHG-derived aerogel stamps (**3**) with contact elements (**4**). a) As-prepared DNHG-derived aerogel stamp. b), c) DNHG-derived aerogel stamp after infiltration with precursor solution (**7**), solvothermal capillary stamping, disassembly of stamping device (**1**) and detachment from ITO substrate (**5**). b) Top view and c) cross-section.

## DNHG-derived aerogel stamps

The DNHGs (Figure 3e) were converted to DNHG-derived aerogels by a two-step solvent exchange procedure from water to methanol to *n*-hexane followed by subcritical drying under ambient conditions. The first solvent exchange from water to methanol also results in the removal of residual reactants such as ethylene glycol and of unreacted acrylamide. The polyacrylamide network tends to minimize its surface area exposed to the non-solvent methanol[48] so that the polyacrylamide precipitates onto the silica network. It is reasonable to assume that the polyacrylamide preferentially agglomerates in the joints between neighboring silica particles. Hence, a tough shell of polyacrylamide encapsulates the brittle silica network. While the tough polyacrylamide network reinforces the stiff silica network, the stiff silica network fixates the tough polyacrylamide network. The second solvent exchange involves the replacement of methanol by *n*-hexane. The lower surface tension of *n*-hexane further reduces the Laplace pressure across the menisci that form in the mesopores during subcritical evaporation of the *n*-hexane under ambient conditions. The DNHG resists the stress occurring during evaporation and undergoes only minor shrinkage, while the structural features of the DNHG are conserved. The DNHG-derived aerogel obtained in this way is a variation of class I hybrid composite aerogels,[49] since the reinforcing polyacrylamide network and the silica network are physically entangled. Even though the silica network and the polyacrylamide network are not covalently connected with each other, they cannot be separated without cleavage of covalent bounds. The DNHG-derived aerogel has a mean pore diameter of 18 nm as well as a total pore volume of 34 $cm^3/g$ (Figures S14 and S15) and combines a low density of 0.5 g * $cm^{-3}$ with an excellent compressive strength of 53.57 MPa ± 7.71 MPa (average of five measurements on different specimens) (Figure S16).

DNHG-derived aerogel stamps with topographically patterned contact surfaces can easily be obtained by carrying out the DNHG synthesis in contact with any kind of mold. The mechanical robustness of the DNHG-derived aerogels should allow the realization of even filigree arbitrary surface structures. For the stamping of ink, the contact surfaces of the topographic stamp features contacting the receiving substrate should exhibit sufficient surface porosity for ink transfer. Considering the spongy pore morphology of the DNHG-derived aerogel stamps used here and their mean pore diameter of 18 nm, we estimate the minimum printable feature size to ~100 nm. Here we used macroporous silicon[36, 37] containing macropores with a diameter of 1 μm and a depth of 730 nm arranged in hexagonal arrays with a pore-center-to-pore-center distance of 1.5 μm as mold. The macroporous silicon was selected because it is commercially available and because its feature sizes match those of 2D photonic crystals with bandgaps in the infrared range.[50] The DNHG-derived aerogel stamps obtained in this way consisted of a monolithic body, which had a height of 13 mm and a topographically patterned contact surface extending 1 $cm^2$. The topographic patterns of the contact surfaces consisted of arrays of rod-like contact elements with a height of ~600 nm and a diameter of 900 nm arranged in a hexagonal lattice with a lattice constant of 1.5 μm (Figure 4a). The shrinkage in the course of the transition from a DNHG in contact with a macroporous silicon mold to a DNHG-derived aerogel stamp does not affect the lattice constant of the arrays of the contact elements because their positions are fixed by the positions of the macropores of the macroporous silicon molds.

**Arrays of submicron nickel ferrite dots by solvothermal capillary stamping**

Solvothermal capillary stamping (Figures 1 and 2) was carried out using a specifically constructed stamping device that consisted of two steel cylinders; an upper part and a lower part. The DNHG-derived aerogel stamps were glued on the upper part. Indium tin oxide (ITO) substrates surface-modified with the silane perfluorodecyltrichlorosilane (FDTS) were placed on the lower part. The upper part and the lower part of the stamping device were assembled in such a way that the contact elements of the DNHG-derived aerogel stamps contacted the FDTS-coated ITO substrates with a contact pressure of 3.9 kN/m$^2$ (Figures 1a and 2a). The assembled stamping device was inserted into a cylindrical PTFE vessel (Figure 2b). An ethanolic precursor solution containing nickel(II)-acetylacetonate and iron(III)-acetylacetonate was deposited through the opening of the PTFE vessel and the central hole in the surface of the upper part of the stamping device onto the DNHG-derived aerogel stamp (Figures 1b and 2c). The precursor solution invaded the DNHG-derived aerogel stamp and filled its mesopore network up to the tips of contact elements. The filling level of the DNHG-derived aerogel stamps could be assessed with the naked eye owing to the brown color of the precursor solution. Figure S17a shows a photograph of a cross-section of a colorless as-prepared DNHG-derived aerogel stamp, whereas Figure S17b shows a photograph of a cross-section of a DNHG-derived aerogel stamp after infiltration with precursor solution. It is obvious that the precursor solution infiltrated the entire volume of ~1.3 cm$^3$ of the DNHG-derived aerogel stamp.

The contact angle of the ethanolic precursor solution on the FDTS-coated rough and grainy ITO substrates under ambient conditions amounted to 52.4° ± 1.5°. Therefore, the precursor solution did not spread on the FDTS-coated ITO substrates. Instead, liquid bridges of the precursor solution connecting the tips of the contact elements of the DNHG-derived aerogel stamps and the surface of the FDTS-coated ITO substrates formed (cf. Figure 1b). The PTFE vessel containing the stamping device including the DNHG-derived aerogel stamp, the FDTS-coated ITO substrate and the precursor solution was inserted into a steel autoclave (Figure 2d,e) with a free volume of ~29.15 mL available to the fluid phases. The solvothermal conversion yielding nickel ferrite was carried out at 413 K for 48 h. We approximated the solvothermal reaction conditions by considering the phase behavior of pure ethanol, which we estimated by the van der Waals equation of state using van der Waals parameters $a$ =12.18 L$^2$ bar/mol$^2$ and $b$ = 0.08407 L/mol.[51] Assuming that 6 mL (~0.1 mol) liquid ethanol is applied to the stamping device with a free volume of 29.15 mL, at 293.15 K the ethanol forms coexisting liquid and gaseous phases at a pressure of 4.3 bar. The mole fraction of the liquid phase amounts to 98 % and that of the gas phase to 2 %. At the reaction temperature of 413 K, liquid and vapor phases coexist at a pressure of 24.6 bar. The mole fractions of the liquid and vapor phases amount to 84 % and 16 %. Thus, under the solvothermal reaction conditions applied here precursor solution and an ethanol-rich vapor phase coexist. However, prior to the supply of precursor solution, no ethanol at all is present in the volume between upper part and lower part of the stamping device. Hence, the ethanol-rich vapor phase is solely obtained by evaporation of ethanol, which can, in principle, occur at the mesopore openings of the DNHG-derived aerogel stamp or at the liquid bridges between the contact elements of the DNHG-derived aerogel stamp and the ITO substrate. However, the concave menisci and the negative Laplace pressure of the precursor solution at the mesopore openings impedes transfer of solvent molecules from the liquid phase to the vapor phase. The liquid bridges do not only have a large exposed liquid surface. Their concave curvature normal to ITO substrate is smaller than the concave curvatures of the menisci at the mesopore openings and impedes evaporation to lesser extent. On the other hand, the curvature of the liquid bridges parallel to the ITO substrate is convex. This convex curvature component will also enhance evaporation, considering that the vapor pressure resulting from liquid surfaces with convex curvature is enhanced as compared to that resulting from plane surfaces, as quantified by the Kelvin equation.[52] Therefore, it is reasonable to assume that the ethanol preferentially evaporates from the liquid bridges into the volume between the upper part and the lower part of the stamping device.

The applied volume of the precursor solution exceeds the volume of the DNHG-derived aerogel stamp by a factor of more than 5 so that the DNHG-derived aerogel stamp is always in contact with a bulk reservoir of the precursor solution located above the upper part of the stamping device. Evaporation drags new precursor solution from the interior of the DNHG-derived aerogel stamp into the liquid bridges. The interior of the DNHG-derived aerogel stamp is in turn refilled from the above-mentioned bulk reservoir of the precursor solution. The nonvolatile nickel ferrite precursors thus enrich in the liquid bridges and are then converted to the solvothermal reaction product nickel ferrite (Figure 1c). We assume that this precursor enrichment mechanism is the main reason for the formation of three-dimensional submicron nickel ferrite dots in place of the liquid bridges. After completion of the solvothermal capillary stamping procedure, removal of the stamping device from the autoclave and the PTFE vessel, disassembly of the stamping device and detachment from the ITO substrate, the DNHG-derived aerogel stamps did not show any damage. While the DNHG-derived aerogel stamps are apparently partially filled with the reaction product nickel ferrite, the contact elements at their contact surface remained intact (Figure 4b,c). The DNHG-derived aerogel stamps could be reused after ultrasonication in ethanol for 30 minutes.

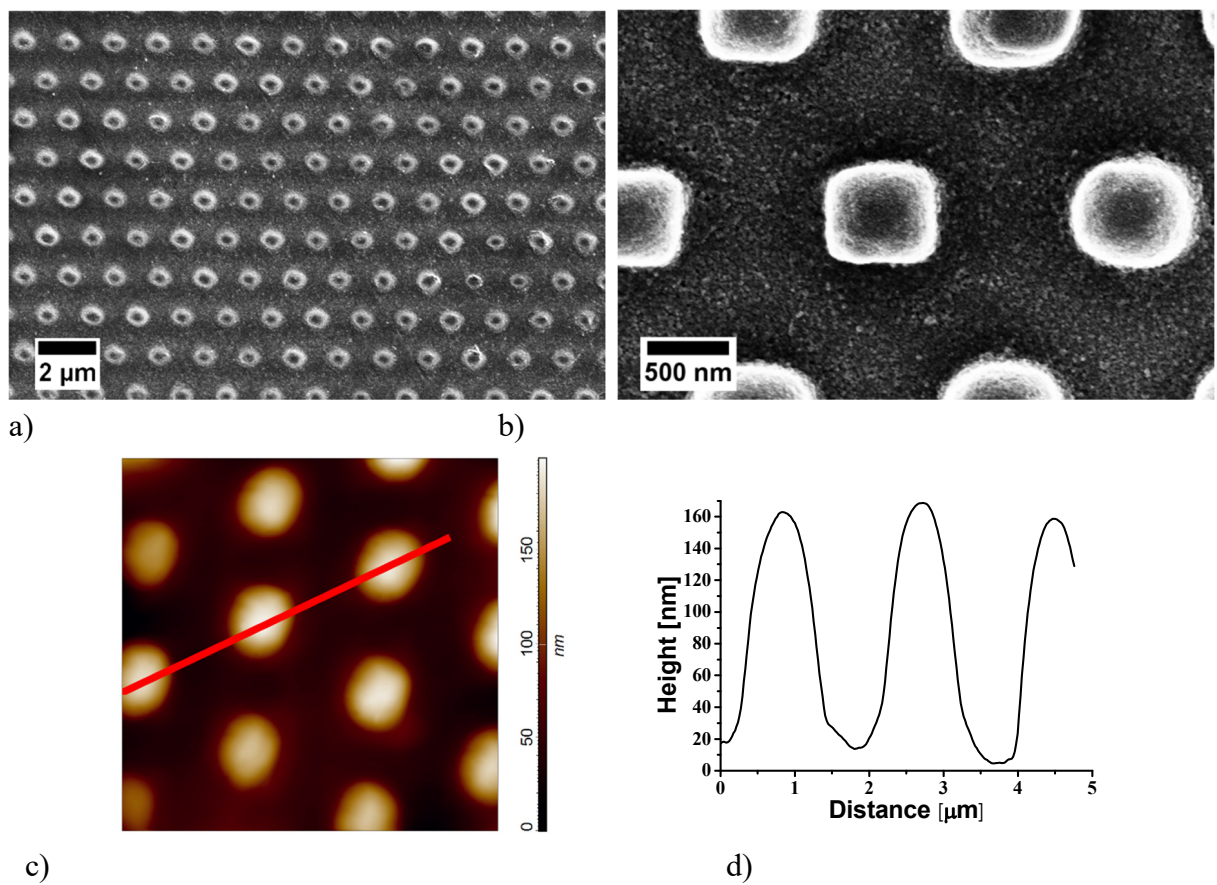

**Figure 5.** Arrays of submicron NiFe$_2$O$_4$ dots on ITO substrates. a) Large-field SEM top-view image. b) SEM top view image at higher magnification. c) AFM topography image (the image field extends 5 x 5 µm$^2$) and d) topographic height profile along the red line in panel c).

**Properties of submicron nickel ferrite dot arrays**

Solvothermal capillary stamping yielded ITO substrates patterned with ordered monolayers of submicron nickel ferrite dots (cf. Figure 1d) typically extending 1 cm$^2$ – corresponding to the area of the contact surface of the DNHG-derived aerogel stamps. The nearest-neighbor distance between the submicron nickel ferrite dots amounted to 1.5 µm and corresponds to the nearest-neighbor distance between the contact elements of the DNHG-derived aerogel stamps. The submicron nickel ferrite dots had a diameter of ~700 nm (Figure 5a,b) and a height of ~150 nm (Figure 5c,d). X-ray photoelectron spectroscopy (XPS) was deployed to evaluate whether the FDTS coating on the ITO substrates was still intact after the solvothermal synthesis (Figure S18). An XP spectrum of an FDTS-coated ITO substrate shows the fluorine 1s peak at 686.43 eV indicating the presence of FDTS. However, after the solvothermal synthesis fluorine was no longer found. Moreover, after the solvothermal synthesis the precursor solution spread on the ITO substrates. These observations corroborate the notion that the solvothermal treatment resulted in the destruction of the FDTS coating.

X-ray powder diffractometry confirmed that any material that could be scraped off from the ITO substrates consisted of nickel ferrite. The diffractogram obtained in this way (Figure 6a) exhibited the characteristic reflections of cubic $NiFe_2O_4$ showing inverse spinel structure (space group Fd-3m), such as the (220) reflection at 2θ = 30.295°, the (311) reflection at 2θ = 35.686°, the (511) reflection at 2θ = 57.377° and the (440) reflection at 2θ = 63.015°. We measured the magnetic moment of an array of submicron nickel ferrite dots on an ITO substrate as function of the external magnetic field by vibrating sample magnetometry (VSM) at 300 K. While bulk nickel ferrite shows ferrimagnetic behavior, the submicron nickel ferrite dots are superparamagnetic with vanishing magnetization hysteresis and vanishing remnant magnetization. The magnetic moment per sample area approached its saturation value of 1.1•10$^{-3}$ emu/cm$^2$ at an external magnetic field of ± 10000 G (Figure 6b). Rescaling the magnetic moment per sample area to the nickel ferrite volume yielded a saturation magnetization of 290 emu/cm$^3$, which is somewhat lower than the saturation magnetization of bulk nickel ferrite amounting to 330 emu/cm$^3$.[53] To approximate the total nickel ferrite volume, we assumed that the submicron nickel ferrite dots were cuboids having bottom square faces with edge lengths of 700 nm and heights of 150 nm, which are arranged in a hexagonal lattice with a lattice constant of 1.5 µm. In this way, we estimated the effective nickel ferrite layer thickness to 38 nm.

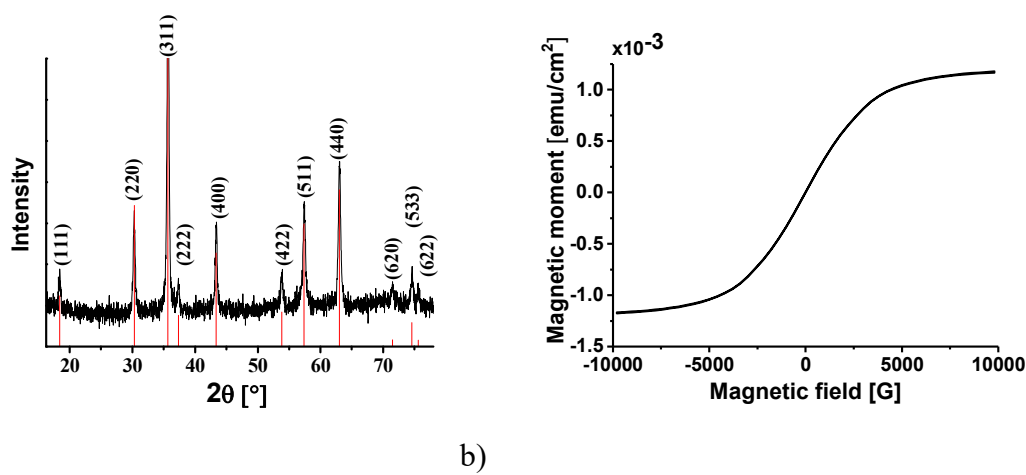

a)                                                   b)

**Figure 6.** Characterization of submicron nickel ferrite dots obtained by solvothermal capillary stamping. a) XRD pattern of a powder of submicron nickel ferrite dots scrapped off from ITO substrates. The red lines indicate characteristic reflections of cubic $NiFe_2O_4$[54] according to PDF cart 01-074-2081 (Inorganic Crystal Structure Database). b) Magnetic moment per sample area of submicron nickel ferrite dots on an ITO substrate extending 1 cm$^2$ as function of an external magnetic field measured by VSM at 300 K.

We used ordered monolayers of submicron nickel ferrite dots on ITO substrates to demonstrate the preparation of thin $NiFe_2O_3$-Ni hybrid films by orthogonal substrate functionalization. Thus, we additionally electrodeposited ~100 nm thick nickel layers on ITO substrates modified with submicron nickel ferrite dots. The nickel was deposited onto the exposed ITO areas surrounding the submicron nickel ferrite dots (Figure 7a,b). In this way, $NiFe_2O_3$-Ni hybrid films consisting of an ordered monolayer of submicron nickel ferrite dots surrounded by a continuous nickel film with a thickness of ~100 nm were obtained (Figure 7c). We further probed the elemental distribution in $NiFe_2O_3$-Ni hybrid films by energy-dispersive X-ray (EDX) spectroscopy. It is challenging to map peaks representing the elements exclusively present in the nickel ferrite dots; oxygen is also contained in the ITO substrates, and the iron peaks were overlapped by much stronger nickel peaks. However, the number of nickel atoms per volume in the nickel ferrite dots is lower than in the surrounding pure nickel films. Therefore, in maps of the intensity of the Ni $L\alpha_{1,2}$ peak the positions of the nickel ferrite dots are apparent as areas in which nickel is depleted (Figure 7d). The dependence of the magnetic moment per sample area of the $NiFe_2O_3$-Ni hybrid film and the pure nickel film on an external magnetic field measured by VSM is displayed in Figure 7e. The saturation values of the magnetic moment per sample area lying in the range from $3.3 \times 10^{-3}$ emu/cm$^2$ to $3.4 \times 10^{-3}$ emu/cm$^2$ were similar for the $NiFe_2O_3$-Ni hybrid film and the pure nickel film. Rescaling the magnetic moment per sample area to the approximated sample volumes yielded saturation magnetizations of 350 emu/cm$^3$ for the pure nickel film and of 290 emu/cm$^3$ for the $NiFe_2O_3$-Ni hybrid film. The volume of the latter was estimated by adding the effective nickel ferrite layer thickness of 38 nm obtained as described above to the effective thickness of the nickel layer amounting to 75 nm. The effective thickness of the nickel layer was approximated by assuming that a 100 nm thick nickel layer was electrodeposited onto the exposed areas of the ITO substrate not covered by nickel ferrite dots. The VSM curve of the $NiFe_2O_3$-Ni hybrid film showed a more pronounced magnetization hysteresis than the VSM curve of the pure nickel film. While the pure nickel film reached the saturation magnetization already at external magnetic field strengths of ±200 G, the $NiFe_2O_3$-Ni hybrid film approached the saturation magnetization only at external field strengths of ±1300 G. The coercive field strengths increased from ±30 G for the pure nickel film to ±90 G for the $NiFe_2O_3$-Ni hybrid film. The remnant magnetic moment per sample area decreased from $3.1 \times 10^{-3}$ emu/cm$^2$ for the pure nickel film to $1.5 \times 10^{-3}$ emu/cm$^2$ for the $NiFe_2O_3$-Ni hybrid film. Rescaling to the estimated sample volumes yielded remnant magnetizations of 310 emu/cm$^3$ for the pure nickel film and of 133 emu/cm$^3$ for the $NiFe_2O_3$-Ni hybrid film. The squareness $M_R/M_S$ of the magnetization hysteresis loops (ratio of the remnant magnetization $M_R$ and the saturation magnetization $M_S$) amounted to ~0.91 for the pure nickel film and to ~0.45 for the $NiFe_2O_3$-Ni hybrid film. Hence, the incorporation of ordered monolayers of nickel ferrite dots into a thin ferromagnetic nickel layers results in magnetic hardening of the latter.

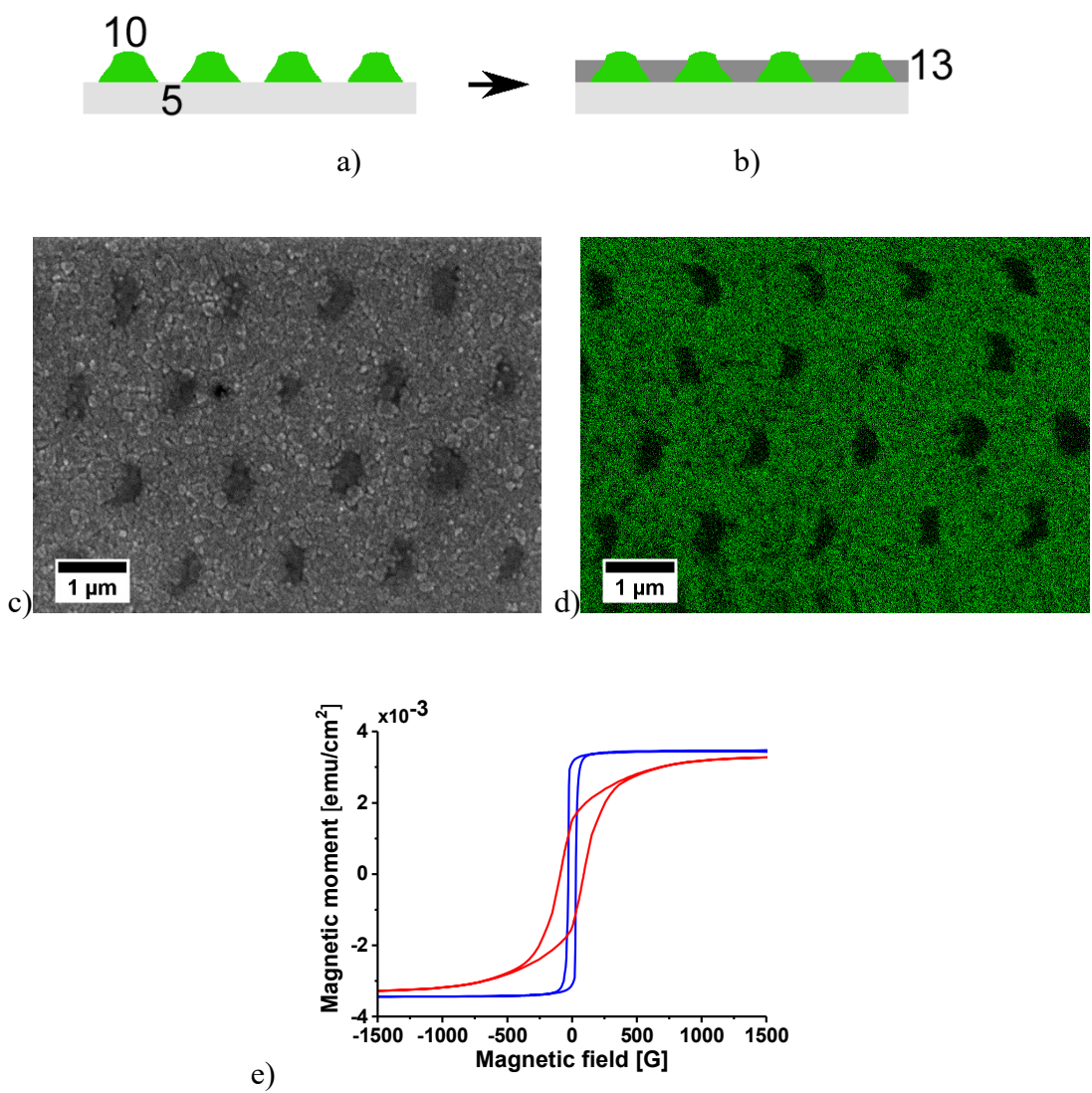

**Figure 7.** Thin NiFe$_2$O$_3$-Ni hybrid films. a) An ITO substrate (**5**) functionalized with an ordered monolayer of submicron nickel ferrite dots (**10**) is b) subjected to orthogonal functionalization by electrodeposition of nickel. Thus, NiFe$_2$O$_3$-Ni hybrid films (**13**) consisting of ordered monolayers of submicron nickel ferrite dots (**10**) surrounded by ~100 nm thick electrodeposited nickel films are obtained on ITO substrates (**5**). c) SEM image of a NiFe$_2$O$_3$-Ni hybrid film (**13**). d) EDX map of the Ni L$\alpha_{1,2}$ peak at 0.85 eV of a NiFe$_2$O$_3$-Ni hybrid film (**13**). Nickel ferrite dots (**10**) are located in areas where the nickel is depleted. e) Magnetic moments per sample area of a NiFe$_2$O$_3$-Ni hybrid film (**13**) (red) on an ITO substrate (**5**) and a ~100 nm thick continuous nickel film electrodeposited on an FDTS-coated ITO substrate (**5**) (blue) as function of an external magnetic field measured by VSM. The samples extended 1 cm$^2$. The magnetization measurements were carried out at 300 K.

## Conclusions

So far, solvothermal syntheses and state-of-the-art lithography have by and large remained incompatible. Thus, the preparation of patterned thin films consisting of solvothermal synthesis products has remained challenging. We have demonstrated capillary stamping in autoclaves under solvothermal conditions enabling stamp-guided conversion of precursors into solvothermal reaction products. The aerogel stamps used for this purpose were derived from a double-network hydrogel by solvent exchange and drying under subcritical ambient conditions. Precursor solutions imbibe the DNHG-derived aerogel stamps. Thus, liquid bridges form between the stamps' contact elements and the substrates to be patterned. Preferential evaporation of the solvent from the liquid bridges drags more precursor solution into the liquid bridges, where the precursors are enriched. Liquid-bridge guided solvothermal syntheses eventually yield submicron dots of the solvothermal reaction products in place of the liquid bridges. Therefore, solvothermal capillary stamping may yield device components comprising complex thin-layer architectures of solvothermal reaction products formed on functional substrates. Problems related to alternative sol-gel routes (high-temperature calcination steps) and direct serial ballistic deposition of the target materials (insufficient adhesion) are overcome. As example, we generated arrays of submicron nickel ferrite dots having heights of ~150 nm on ITO substrates. Subsequent orthogonal substrate functionalization by electrodeposition of nickel onto the exposed substrate areas surrounding the submicron nickel ferrite dots yielded thin nickel ferrite-nickel hybrid films consisting of a monolayer of submicron nickel ferrite dots surrounded by a continuous nickel layer. The submicron nickel ferrite dots halved the remanence of the ferromagnetic nickel film, while the saturation value of the magnetic moment per area remained by and large constant. Remanence engineering of thin ferromagnetic layers may help customize miniaturized transformers, positioning systems and electromechanical systems such as nanorelays. Solvothermal capillary stamping may pave the way for complex functional thin-film configurations so far predominantly accessible by combinations of layer transfer techniques, such as wafer bonding or ion slicing, and subtractive lithography. Orthogonal functionalization of substrates patterned by solvothermal capillary stamping may yield functional hybrid layers, in which the properties of the solvothermal reaction product and the second component are either synergistic or complementary.

## ASSOCIATED CONTENT

**Supporting Information**. NMR characterization of $Si(OC_2H_4OH)_4$; cross-sectional SEM image of an electrodeposited Ni layer; mechanical characterization of polyacrylamide hydrogels and DNHGs; characterization of DNHG-derived aerogels by nitrogen sorption measurements, SEM and stress-strain measurements; photographs of non-infiltrated and infiltrated DNHG-derived stamps; XPS characterization of substrate surfaces. This material is available free of charge via the Internet at http://pubs.acs.org."

## AUTHOR INFORMATION

**Corresponding Author**

* Martin Steinhart, Institut für Chemie neuer Materialien and CellNanOs, Universität Osnabrück, Barbarastr. 7, 49076 Osnabrück, Germany; https://orcid.org/0000-0002-5241-8498; Email: martin.steinhart@uos.de

**Author Contributions**

The manuscript was written through contributions of all authors. / All authors have given approval to the final version of the manuscript.

‡ These authors contributed equally.

**Funding Sources**

The authors thank the European Research Council (ERC-CoG-2014, Project 646742 INCANA) for funding.

**Notes**

The authors declare no competing financial interest.